\RequirePackage{arydshln}
\documentclass[floatfix,aps,twocolumn,nofootinbib,superscriptaddress,preprintnumbers,pra,10pt]{revtex4-1}

\usepackage[utf8]{inputenc}
\usepackage{float}
\usepackage{colortbl}
\usepackage{amsmath,amssymb,amsfonts,bm,bbm,slashed}
\usepackage{dsfont} 
\usepackage{hyperref}
\usepackage{graphicx}
\usepackage{enumitem}
\usepackage{arydshln}
\usepackage{mathtools}
\usepackage{bbold}
\usepackage{braket}
\usepackage{multirow}
\usepackage{soul}
\usepackage{xcolor}
\usepackage{soul}

\usepackage{lipsum}  

\usepackage{pifont}
\newcommand{\xmark}{\ding{55}}%

\makeatletter
\g@addto@macro\bfseries{\boldmath}
\makeatother

\newcommand{\gsim}{\lower.7ex\hbox{$\;\stackrel{\textstyle>}{\sim}\;$}}
\newcommand{\lsim}{\lower.7ex\hbox{$\;\stackrel{\textstyle<}{\sim}\;$}}

\newcommand{\bea}{\begin{eqnarray}}
\newcommand{\eea}{\end{eqnarray}}

\allowdisplaybreaks

\begin{document}

\preprint{ }

\title{On the single leptoquark solutions to the $B$-physics anomalies}
 
\author{Andrei Angelescu}
\email{andrei.angelescu@mpi-hd.mpg.de}
\affiliation{Max-Planck-Institut für Kernphysik, Saupfercheckweg 1, 69117 Heidelberg, Germany}
\author{Damir Be\v{c}irevi\'c}
\email{damir.becirevic@ijclab.in2p3.fr}
\affiliation{IJCLab, Pôle Théorie (Bât.~210), CNRS/IN2P3 et Université Paris-Saclay,
91405 Orsay, France}
\author{Darius A.~Faroughy}
\email{faroughy@physik.uzh.ch}
\affiliation{Physik-Institut, Universit\"at Z\"urich, CH-8057 Z\"urich, Switzerland}
\author{Florentin Jaffredo}
\email{florentin.jaffredo@ijclab.in2p3.fr}
\affiliation{IJCLab, Pôle Théorie (Bât.~210), CNRS/IN2P3 et Université Paris-Saclay,
91405 Orsay, France}
\author{Olcyr Sumensari}
\email{olcyr.sumensari@ijclab.in2p3.fr}
\affiliation{IJCLab, Pôle Théorie (Bât.~210), CNRS/IN2P3 et Université Paris-Saclay,
91405 Orsay, France}

\begin{abstract}
\vspace{5mm}
We revisit the possibilities of accommodating the experimental indications of the lepton flavor universality violation in $b$-hadron decays in the minimal scenarios 
in which the Standard Model is extended by the presence of a single $\mathcal{O}(1\,\mathrm{TeV})$ leptoquark state. 
To do so we combine the most recent low energy flavor physics constraints, including $R_{K^{(\ast)}}^\mathrm{exp}$ and $R_{D^{(\ast)}}^\mathrm{exp}$, and combine them 
with the bounds on the leptoquark masses and their couplings to quarks and leptons as inferred from the direct searches at the LHC and the studies of the 
large $p_T$ tails of the $pp\to \ell\ell$ differential cross section. 
We find that none of the scalar leptoquarks of $m_\mathrm{LQ} \simeq 1\div 2$~TeV can accommodate the $B$-anomalies alone. Only the vector leptoquark, known as $U_1$, can provide a viable solution which, in the minimal setup, provides an interesting prediction, i.e.  a lower bound to the lepton flavor violating $b\to s\mu^\pm\tau^\mp$ decay modes, such as  $\mathcal{B}(B\to K\mu\tau) \gtrsim 0.7\times 10^{-7}$.
\vspace{3mm}
\end{abstract}

\maketitle

\section{Introduction}\label{sec:intro}


In Ref.~\cite{Angelescu:2018tyl} we made a comprehensive phenomenological analysis of the new physics (NP) scenarios  
in which the Standard Model (SM) is extended minimally by a single $\mathcal{O}(1\,\mathrm{TeV})$ leptoquark state. The purpose of that study was to examine which one of the known 
leptoquarks can be made compatible with the experimental indications of the lepton flavor universality violation (LFUV), as inferred from the decays of $b$-flavored hadrons, and be consistent with many other flavor observables, as well as with the direct and indirect NP searches at the LHC. 
Since the publication of that study several new measurements appeared, and some of the theoretical estimates have been improved. More specifically: 
\begin{itemize} 
\item LHCb collaboration presented their new result for $R_K$~\cite{1852846} which now, combined with their previous data, amounts to
\begin{equation}\label{newRK}
R_K^{[1.1,6]} =0.847 \pm 0.042\,,
\end{equation}
which is $3.1\sigma$ lower than predicted in the SM, $R_K^{[1,6]} = 1.00(1)$~\cite{Bordone:2016gaq}.~\footnote{We combined the errors in quadrature before symmetrizing them.}
We remind the reader that the ratios
\begin{equation}
R_{K^{(\ast)}}^{[q_1^2, q_2^2]} =  \dfrac{\mathcal{B}'(B\to K^{(\ast)} \mu\mu)}{\mathcal{B}'(B\to K^{(\ast)} ee)}  \,,
\label{eq:RK_definition}
\end{equation}
are defined in terms of partial branching fractions ($\mathcal{B}'$), corresponding to a  
conveniently chosen interval $q_1^2\leq q^2\leq  q_2^2$ as to stay away from the prominent $c\bar c$-resonances. 
In this paper, in addition to the value~\eqref{newRK}, we will also use~\cite{Aaij:2017vbb}
\begin{equation}
\;\;\;\;R_{K^\ast}^{[0.045,1.1]} =0.68 \pm 0.10\,, \quad R_{K^\ast}^{[1.1,6]} =0.71 \pm 0.10\,.
\end{equation}
Notice that a hint of LFUV has also been observed in the decay of $\Lambda_b$~\cite{Aaij:2019bzx}.

\item The experimental value of $\mathcal{B}(B_s\to\mu\mu)$ has been recently updated to~\cite{CMS:2020rox}
\begin{equation}
\mathcal{B}(B_s\to\mu\mu) = (2.70 \pm 0.36) \times 10^{-9} \,,
\label{eq:Bsmumu-new}
\end{equation}
to which we include the most recent update of the LHCb result $\mathcal{B}(B_s\to\mu\mu) = (3.09^{+0.48}_{-0.44}) \times 10^{-9} $~\cite{LHCbNEW}, and by using the prescription of Ref.~\cite{Barlow:2004wg} to build the likelihood functions, the new average value is
\begin{equation}
\mathcal{B}(B_s\to\mu\mu) = (2.85 \pm 0.33) \times 10^{-9} \,,
\end{equation}
thus a little over $2 \sigma$ lower than predicted in the SM, $\mathcal{B}(B_s\to\mu\mu) =3.66(14)\times 10^{-9}$~\cite{Beneke:2019slt}.

\item Experimental indications of LFUV have also been observed in the $b\to c\ell \bar \nu_\ell$ decays, and more specifically in
\begin{equation}
R_{D^{(\ast)}} = \left. \dfrac{\mathcal{B}(B\to D^{(\ast)} \tau\bar{\nu})}{\mathcal{B}(B\to D^{(\ast)} l \bar{\nu})}\right|_{l\in \{e,\mu\}}.
\label{eq:RD_definition}
\end{equation}
Recent measurements by Belle~\cite{Abdesselam:2019dgh}, lead to the new averages~\cite{Amhis:2019ckw},
\begin{equation}
R_{D} =0.340\pm 0.030\,, \quad R_{D^\ast} =0.295\pm 0.014\,,
\end{equation}
which are, due to experimental correlations, about $\approx 3 \sigma$ larger than predicted in the SM (see~\cite{Amhis:2019ckw} and references therein), 
\begin{equation}
 R_D^\mathrm{SM} = 0.293\pm0.008\,,\quad R_{D^\ast}^\mathrm{SM} =0.257\pm 0.003\,. 
\end{equation}
A similar deviation, but with less competitive experimental uncertainties, has been observed in a similar $R_{J/\psi}$ ratio~\cite{Aaij:2017tyk}.

\item Direct searches for the leptoquark states, either via the pair production of leptoquarks or through a study of the high $p_T$ tails of the differential cross section of $pp\to \ell\ell$, have been significantly improved, resulting in ever more stringent bounds on masses and (Yukawa) couplings relevant to the results presented here.  
\end{itemize}

In the following we will use the above experimental improvements, combine them with theoretical expressions used in Ref.~\cite{Angelescu:2018tyl} and references therein, or with the improved expressions which will be properly referred to in the body of this letter organized as follows: In Sec.~\ref{sec:eft} we update the effective field theory (EFT) analysis of the transitions $b\to s\mu\mu$ and $b\to c\tau\bar{\nu}$ to determine the effective coefficients that can accommodate the latest experimental results for $R_{K^{(\ast)}}$ and $R_{D^{(\ast)}}$. In Sec.~\ref{sec:lqs}, we remind the reader of the leptoquark (LQ) states that can induce the viable effective operators. In Sec.~\ref{sec:lhc}, we derive updated limits on the LQ mass and couplings by using the most recent LHC results at high-$p_T$. In Sec.~\ref{sec:which}, we combine the low and high-energy constraints to determine which LQs can accommodate the LFU discrepancies. Our findings are summarized in Sec.~\ref{sec:conclusions}.

\section{Effective field theory}\label{sec:eft}

\subsection{$R_K$ and $R_{K^\ast}$}

The effective Lagrangian for a generic exclusive decay based on $b\to s \ell_1^- \ell_2^+$, with $\ell_{1,2}\in\{e,\mu,\tau\}$ can be written as 
\begin{equation}
\label{eq:lagrangian-bsll}
\begin{split}
  \mathcal{L}_{\mathrm{nc}} \supset \frac{4
    G_F}{\sqrt{2}}V_{tb} V_{ts}^\ast &\sum_{i}
  C_i\,\mathcal{O}_i
+\mathrm{h.c.}\,,
\end{split}
\end{equation}

\noindent where the effective couplings (Wilson coefficients) $C_i \equiv C_i (\mu)$ and the operators $\mathcal{O}_{i}\equiv\mathcal{O}_{i}(\mu)$ are defined at the scale $\mu$. The  operators relevant to this study are
\begin{align}
\label{eq:C_LFV}
\begin{split}
\mathcal{O}_{9}^{\ell_1\ell_2}
  &=\frac{e^2}{(4\pi)^2}(\bar{s}\gamma_\mu P_{L}
    b)(\bar{\ell}_1\gamma^\mu\ell_2)\,, \\
    \mathcal{O}_{10}^{\ell_1\ell_2} &=
    \frac{e^2}{(4\pi)^2}(\bar{s}\gamma_\mu P_{L}
    b)(\bar{\ell}_1\gamma^\mu\gamma^5\ell_2)\,,\\
\mathcal{O}_{S}^{\ell_1\ell_2} &=
  \frac{e^2}{(4\pi)^2}(\bar{s} P_{R} b)(\bar{\ell}_1 \ell_2)\,,\\
\mathcal{O}_{P}^{\ell_1\ell_2} &=
  \frac{e^2}{(4\pi)^2}(\bar{s} P_{R} b)(\bar{\ell}_1 \gamma^5 \ell_2)\,,\\
\end{split}
\end{align}
in addition to the chirality flipped ones, $\mathcal{O}_i^\prime$, obtained from $\mathcal{O}_i$ by replacing $P_L\leftrightarrow P_R$. The effect of operators $\mathcal{O}_{1-6}$ is included in the redefinition of the effective Wilson coefficients $C_{7,9}$. In what follows we ignore the electromagnetic dipole operators $\mathcal{O}_{7}^{(\prime)}$ since they do not play a significant role in describing the effects of LFUV. Starting from Eq.~\eqref{eq:lagrangian-bsll} it is straightforward to compute the decay rates for $B_s\to \ell_1^-\ell_2^+$, $B\to K^{(\ast)}\ell_1^-\ell_2^+$, and $\Lambda_b\to \Lambda \ell_1^-\ell_2^+$ see e.g.~Refs.~\cite{Becirevic:2016zri,florentin2}. In the following the NP contributions to $b\to s\ell_1^-\ell_2^+$ will be denoted by $\delta C_i^{\ell_1 \ell_2}$.~\footnote{From now on we will drop the electric charges for the LFV modes and denote $\mathcal{B}(B\to K^{(\ast)}\ell_1 \ell_2) =
\mathcal{B}(B\to K^{(\ast)}\ell_1^-\ell_2^+)+\mathcal{B}(B\to K^{(\ast)}\ell_1^+\ell_2^-)$. }

After neglecting the NP couplings to electrons, it has been established that in order to simultaneously accommodate $R_K^\mathrm{exp} < R_K^{\mathrm{SM}}$ and $R_{K^\ast}^\mathrm{exp} < R_{K^\ast}^{\mathrm{SM}}$, the preferred scenarios are those with $\delta C_9^{\mu\mu}<0$, or those in which $\delta C_9^{\mu\mu}=-\delta C_{10}^{\mu\mu}<0$. This conclusion has been corroborated by numerous global analyses of the $b\to s\mu\mu$ observables~\cite{Capdevila:2017bsm}. In this work, we adopt a conservative approach by only taking into account the LFUV ratios ($R_{K}^\mathrm{exp}$, $R_{K^{\ast}}^\mathrm{exp}$) and $\mathcal{B}(B_s\to\mu\mu)^\mathrm{exp}$, the quantities for which the hadronic uncertainties are very small and well under control. Notice that the subpercent precision of the lattice QCD determination of the decay constant entering $\mathcal{B}(B_s\to\mu\mu)^\mathrm{exp}$ is also a very recent achievement, $f_{B_s}= 230.3\pm 1.3$~MeV~\cite{Aoki:2019cca}.

\begin{figure}[t!]
\centering
\includegraphics[width=1.\linewidth]{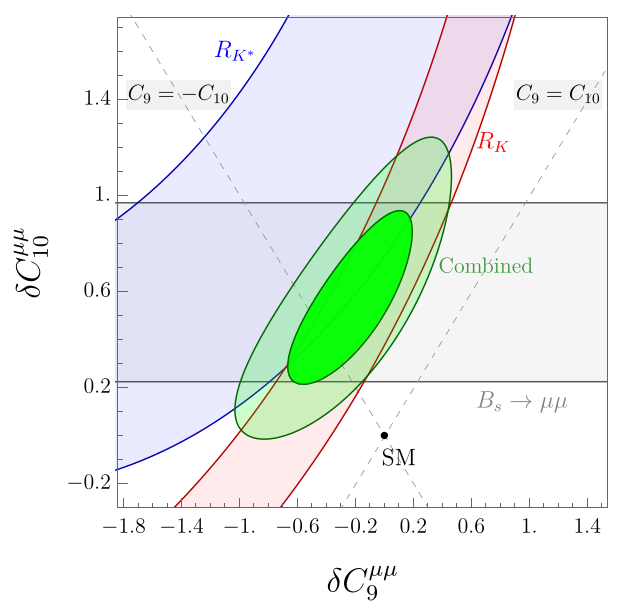}
\caption{\small \sl Allowed regions in the plane $\delta C_9^{\mu\mu}$ vs.~$\delta C_{10}^{\mu\mu}$ to $1\sigma$ accuracy derived by using $R_K$ (red region), $R_{K^\ast}$ (blue region) and $\mathcal{B}(B_s\to\mu\mu)$ (gray region). Darker (lighter) green regions correspond to the combined fit to $1\sigma$ ($2\sigma$) accuracy. }
\label{fig:c9-c10-fit} 
\end{figure}

The result of our fit is shown in Fig.~\ref{fig:c9-c10-fit} where we see a good agreement among all three observables. Furthermore, we again see that the data are not consistent with the scenario 
$\delta C_9^{\mu\mu} = + \delta C_{10}^{\mu\mu}$, but instead they are consistent with the solution, $\delta C_9^{\mu\mu} = - \delta C_{10}^{\mu\mu}$. By focussing onto the latter, we find 
\begin{align}
\delta C_9^{\mu\mu} = - \delta C_{10}^{\mu\mu} = -0.41\pm 0.09 \,,
\end{align}
which measures the deviation between the measured and the SM predictions of all three observables combined.

\subsection{$R_D$ and $R_{D^\ast}$}

We remind the reader of the most general low-energy EFT describing the $b\to c\ell \bar{\nu}$ decay with operators up to dimension-six,
	\begin{align}
		\label{eq:lagrangian-lep-semilep}
		\mathcal{L}_{\mathrm{cc}} &= -2\sqrt{2}G_F V_{cb} \Big{[}(1+g_{V_L})\,(\bar{c}_{L}\gamma_\mu {b}_{L}) (\bar{\ell}_L\gamma^\mu\nu_L)\nonumber \\
		&+ g_{V_R}\,(\bar{c}_{R}\gamma_\mu {b}_{R}) (\bar{\ell}_L\gamma^\mu\nu_L) +g_{S_R}\,(\bar{c}_{L} b_{R})(\bar{\ell}_R \nu_L) \\
		&+g_{S_L}\,(\bar{c}_{R} b_{L})(\bar{\ell}_R \nu_L)+g_T\,(\bar{c}_R \sigma_{\mu\nu}b_L)(\bar{\ell}_R \sigma^{\mu\nu} \nu_L)\Big{]}+\mathrm{h.c.}\,,\nonumber
	\end{align}
where  the NP couplings, $g_i \equiv g_i (\mu)$, are defined at the renormalization scale which in the following will be taken to be $\mu=m_b$. 
Flavor indices in $g_i$ are omitted for simplicity. 

To determine the allowed values of $g_i$, we assume that NP predominantly contributes to the $b\to c\tau \bar{\nu}$ transition, while being tiny in the case of electron or muon in the final state. In addition to the ratios $R_D$ and $R_{D^\ast}$, an important constraint onto $g_P\equiv g_{S_R}-g_{S_L}$ comes from the $B_c$-meson lifetime~\cite{Alonso:2016oyd}. In that respect, we conservatively impose on the still unknown decay rate to be $\mathcal{B} (B_c\to\tau \bar{\nu})\lesssim  30\%$. That constraint alone already eliminates a possibility of accommodating the $R_{D^{(\ast)}}^\mathrm{exp}$ values by solely relying on the (pseudo)scalar operators~\cite{Alonso:2016oyd}. 

By using the hadronic input collected in Ref.~\cite{Angelescu:2018tyl} we make the one-dimensional fits in which one real effective coupling at a time is allowed to take a non-zero value, $g_i(m_b)$, where $i\in \lbrace V_L, S_R, S_L, T \rbrace$. We also consider two scenarios motivated by the LQ models and defined by the relations $g_{S_L}(\Lambda) = + 4\, g_T(\Lambda)$ and $g_{S_L}(\Lambda) = - 4\, g_T(\Lambda)$ at the scale $\Lambda \approx 1$~TeV. After accounting for the renormalization group running from $\Lambda$ to $m_b$, these relations become $g_{S_L}(m_b) \approx + 8.1\, g_T(m_b)$ and $g_{S_L}(m_b) \approx - 8.5\, g_T(m_b)$, respectively. We quote the allowed $1\sigma$ ranges for $g_{S_L}(m_b)$ in the latter two scenarios, both for real and for purely imaginary values. The results of all these scenarios are presented in Table~\ref{tab:bctaunu-fit}, where we see that only a few scenarios can improve the SM description of $b\to c\tau \bar{\nu}$ data.

\begin{table}[t]
\renewcommand{\arraystretch}{1.95}
\centering
\begin{tabular}{|c|c|c|}
\hline 
Eff.~coeff.  &  ~~$1\sigma$ range~~ & ~$\chi^2_\mathrm{min}/\mathrm{dof}$~ \\ \hline\hline
 $g_{V_L}(m_b)$   &  $0.07\pm 0.02$  & $0.02/1$  \\  
 $g_{S_R}(m_b)$   &  $-0.31\pm 0.05$  &  $5.3/1$ \\  
 $g_{S_L}(m_b)$   &  $0.12\pm 0.06$  & $8.8/1$  \\  
 $g_{T}(m_b)$   &  $-0.03\pm 0.01$  & $3.1/1$  \\  \hline
 $g_{S_L} = +4g_T \in \mathbb{R} $   & $-0.03\pm 0.07$ & $12.5/1$ \\ 
 $g_{S_L} = -4 g_T \in \mathbb{R} $    & $0.16\pm 0.05$ & $2.0/1$ \\ 
 $g_{S_L} = \pm 4  g_T \in i\,\mathbb{R} $    & $0.48\pm 0.08$ & $2.4/1$ \\
 \hline
\end{tabular}
\caption{ \sl \small Low-energy fit to the $b\to c\tau \bar{\nu}$ effective coefficients defined in Eq.~\eqref{eq:lagrangian-lep-semilep} by using $R_D$ and $R_{D^\ast}$, and by imposing that $\mathcal{B}(B_c\to\bar{\tau}\nu)\lesssim 30\%$. For the individual effective coefficients $g_a$, we fix the renormalization scale at $\mu=m_b$. For the remaining scenarios with both $g_{S_L}$ and $g_T$, we impose the conditions $g_{S_L}= \pm 4 g_T$ at $\Lambda =1$~TeV, and provide the allowed range for $g_{S_L}(m_b)$ after accounting for the renormalization-group evolution. The values of $\chi^2_\mathrm{min}$ for each scenario is to be compared to $\chi^2_\mathrm{SM}=12.7$.}
\label{tab:bctaunu-fit} 
\end{table}

\begin{figure}[t!]
\centering
\includegraphics[width=.95\linewidth]{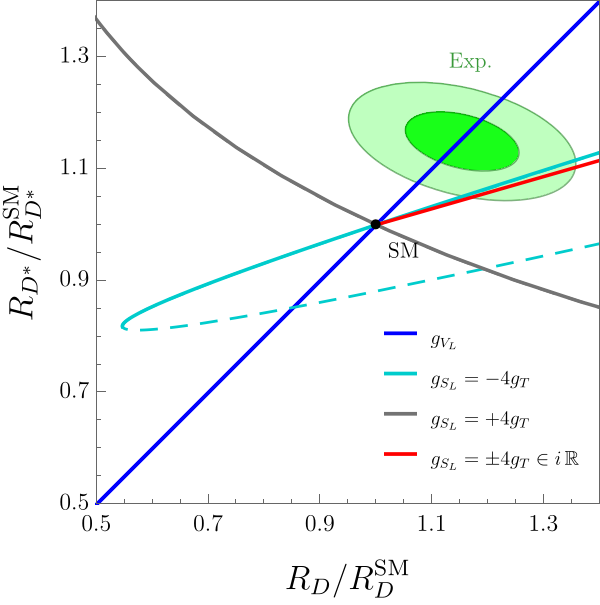}\\
\hfill\includegraphics[width=.975\linewidth]{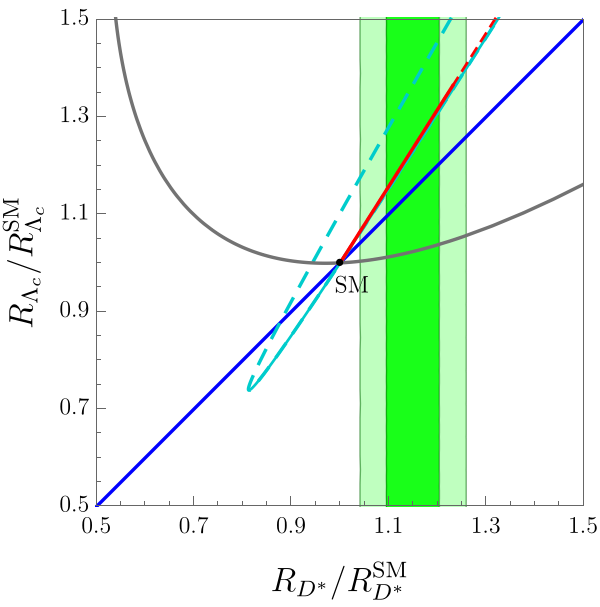} 
\caption{\small \sl Predictions for $R_{D^\ast}/R_{D^\ast}^\mathrm{SM}$ and $R_{\Lambda_c}/R_{\Lambda_c}^\mathrm{SM}$ versus $R_D/R_{D}^\mathrm{SM}$ in several EFT scenarios, see text for details. Current $1\sigma$ ($2\sigma$) experimental constraints are depicted by the darker (lighter) green region. Dashed lines correspond to effective couplings that are in tension with the $\mathcal{B}(B_c\to\tau\nu)<0.3$ constraint.}
\label{fig:rd-rdst-fit}
\end{figure}

In Fig.~\ref{fig:rd-rdst-fit}, we predict the correlation between $R_{D^\ast}/R_{D^\ast}^\mathrm{SM}$ and $R_D/R_D^\mathrm{SM}$ within selected EFT scenarios, and we confront these predictions with the current experimental values for these ratios. In this plot, we also illustrate the results presented in Table~\ref{tab:bctaunu-fit} and confirm that the scenarios with  $g_{V_L}>0$, $g_{S_L}=-4 g_T>0$ and $g_{S_L}=\pm 4 g_T \in i \,\mathbb{R}$ are in good agreement with current data. Furthermore, it becomes clear why the scenario $g_{S_L}=4 g_T\in \mathbb{R}$ is excluded, as it cannot simultaneously explain an excess in both $R_D^\mathrm{exp}$ and $R_{D^\ast}^\mathrm{exp}$. In the same Fig.~\ref{fig:rd-rdst-fit}, we show a similar correlation between $R_{\Lambda_c}/R_{\Lambda_c}^\mathrm{SM}$ and $R_{D^\ast}/R_{D^\ast}^\mathrm{SM}$, which is perhaps more interesting a prediction, since the value of $R_{\Lambda_c} = \mathcal{B}(\Lambda_b\to \Lambda_c \tau\bar{\nu})/\mathcal{B}(\Lambda_b\to \Lambda_c \mu \bar{\nu})$ has not yet been experimentally established, although the early study has been reported in Ref.~\cite{Renaudin}.
Theoretical expressions for $R_{\Lambda_c}$ in a general NP scenario \eqref{eq:lagrangian-lep-semilep} can be found in Ref.~\cite{RLambdac}.

\section{Leptoquarks for $R_{K^{(\ast)}}$ and $R_{D^{(\ast)}}$}\label{sec:lqs}

In this Section we discuss which LQ can be added to the SM in order to accommodate one or both types of the LFUV ratios,  $R_{K^{(\ast)}}$ and $R_{D^{(\ast)}}$. 
We refer the reader to our previous paper~\cite{Angelescu:2018tyl} for a more extensive discussion. We specify each LQ by its SM quantum numbers $({SU(3)}_c,{SU(2)}_L,U(1)_Y)$, where the electric charge, $Q=Y+T_3$, is the sum of the hypercharge $(Y)$ and the third-component of weak isospin $(T_3)$. We neglect the possibility of right-handed neutrinos and we work in the basis with diagonal lepton and down-quark Yukawas, i.e.~with left-handed doublets $Q_i=\big{(}(V^\dagger u_L)_i~d_{Li}\big{)}^T$ and $L_i= \big{(}\nu_{Li}~\ell_{Li}\big{)}^T$, where $V$ stands for the CKM matrix.

\subsection{Scalar leptoquarks}

\begin{itemize}
 \item[$\bullet$] $S_3=(\mathbf{\bar{3}},\mathbf{3},1/3)$\,: The weak triplet of LQs is the only scalar boson that can simultaneously accommodate $R_K^\mathrm{exp}<R_K^\mathrm{SM}$ and $R_{K^\ast}^\mathrm{exp}<R_{K^\ast}^\mathrm{SM}$ at tree level~\cite{Hiller:2014yaa,Dorsner:2017ufx}. The Yukawa Lagrangian of $S_3$ can be written as
\begin{equation}
 \begin{aligned}
  \mathcal{L}_{S_3} =  y_L^{ij}\, \overline{Q^C_{i}} i \tau_2 \big{(}\vec{\tau}\cdot \vec{S_3}\big{)}L_j +\mathrm{h.c.}\,,
 \end{aligned}
\end{equation}
\noindent where $\tau_k$ are the Pauli matrices $(k=1,2,3)$ and $y_{L(R)}^{ij}$ the generic Yukawa couplings with quark (lepton) indices $i(j)$. LQ couplings to diquarks are neglected in order to guarantee the proton stability~\cite{Dorsner:2016wpm}. After integrating out the LQ, we find that the $b\to s\ell_l^-\ell_k^+$ effective coefficients read 
\begin{equation}
 \begin{aligned}
  \delta C_9^{kl} = - \delta C_{10}^{kl} = \dfrac{\pi v^2}{V_{tb} V_{ts}^\ast \alpha_\mathrm{em}} \dfrac{y_L^{bk} \big{(}y_L^{sl}\big{)}^\ast}{m_{S_3}^2}\,,
 \end{aligned}
\end{equation}
which is indeed a pattern that can accommodate $b\to s\mu\mu$ data, cf. Fig.~\ref{fig:c9-c10-fit}. As for the charged current transitions, $b\to c\ell \bar{\nu}_{\ell^\prime}$,  the $S_3$ scenario generates at tree level 
\begin{equation}
 \begin{aligned}
g_{V_L} = - \frac{v^2}{4 V_{cb}}\, \frac{y_L^{b\ell^\prime} (V y_L)_{c\ell} }{ m_{S_3}^2 } \,,
 \end{aligned}
\end{equation}
which is strictly negative if we account for the constraints coming from $B\to K^{(\ast)}\nu \bar{\nu}$ and $\Delta m_{B_s}$~\cite{Angelescu:2018tyl}. Therefore, this scenario is in conflict with results presented in Table~\ref{tab:bctaunu-fit} and it cannot accommodate $R_{D^{(\ast )}}^\mathrm{exp}>R_{D^{(\ast )}}^\mathrm{SM}$ as a small and positive $g_{V_L}$ value is  needed.

 \item[$\bullet$] $S_1=(\mathbf{\bar{3}},\mathbf{1},1/3)$\,: The weak singlet scalar LQ has the peculiarity of contributing to the $b\to c\tau \bar{\nu}$ transition at tree level, but only at loop level to $b\to s\ell\ell$~\cite{Bauer:2015knc}. The $S_1$ Yukawa Lagrangian reads
\begin{equation}
 \begin{aligned}
 \qquad\mathcal{L}_{S_1} =  y_L^{ij}\, \overline{Q^C_{i}} i \tau_2 L_j\,S_1 + y_R^{ij}\, \overline{u_{Ri}^C} \ell_{Rj}\,S_1+\mathrm{h.c.}\,,
 \end{aligned}
\end{equation}
where $y_L$ and $y_R$ are the LQ Yukawa matrices, and we neglect the diquark couplings for the same reason as in the $S_3$ case. The coefficients $C_9^{kl}+C_{10}^{kl}$ and $C_9^{kl}-C_{10}^{kl}$ are generated at one-loop by $y_L$ and $y_R$, respectively, with the relevant expressions provided in Ref.~\cite{Bauer:2015knc}. 
This scenario contributes to the $b\to c\ell \bar{\nu}_{\ell^\prime}$ transitions via, 
\begin{align}\label{eq:gVSS1}
g_{V_L} &= \dfrac{v^2}{4 V_{cb}}\dfrac{y_L^{b\ell^\prime}\big{(}V y_L^\ast \big{)}_{c\ell}}{m_{S_1}^2}\,,\\[0.4em]
g_{S_L} &= - 4 g_T =-\dfrac{v^2}{4 V_{cb}}\dfrac{y_L^{b\ell^\prime}\big{(}y_R^{c\ell}\big{)}^\ast}{m_{S_1}^2}\,,
\end{align}
at the matching scale $\mu=m_{S_1}$. Note, in particular, that both $g_{V_L}$ and $g_{S_L}=- 4 g_T$ can accommodate the observed excesses in $R_D$ and $R_{D^\ast}$, see also Fig.~\ref{fig:rd-rdst-fit}.

 \item[$\bullet$] $R_2=(\mathbf{{3}},\mathbf{2},7/6)$\,: The weak doublet was proposed to separately explain the LFUV effects in the charged~\cite{Sakaki:2013bfa,Becirevic:2018afm} and in the neutral current $B$-decays~\cite{Becirevic:2017jtw}. This is the only scalar LQ that automatically conserves baryon number~\cite{Assad:2017iib}. Its Yukawa Lagrangian writes
\begin{equation}
 \begin{aligned}
 \quad \mathcal{L}_{R_2} = - y_L^{ij}\, \overline{u}_{Ri} R_2 i \tau_2 L_j + y_R^{ij}\, \overline{Q}_i R_2 \ell_{Rj}+\mathrm{h.c.}\,,
 \end{aligned}
\end{equation}
with $y_L$ and $y_R$ being the LQ couplings to fermions. At tree level one gets,
\begin{equation}
\delta C_{9}^{kl} = \delta C_{10}^{kl} \overset{\mathrm{tree}}{=} -\dfrac{\pi v^2}{2 V_{tb} V_{ts}^\ast \alpha_\mathrm{em}} \dfrac{y_R^{sk} \big{(}y_R^{bl}\big{)}^\ast}{m_{R_2}^2}\,,
\end{equation}
a pattern excluded by the observed values of $R_K$ and $R_{K^\ast}$, viz. Fig.~\ref{fig:c9-c10-fit}. If, however, one sets $y_R = 0$, the leading contribution to $b\to s\mu\mu$ arises at one-loop level and the Wilson coefficients verify $\delta C_{9}^{\mu\mu} = -\delta C_{10}^{\mu\mu}<0$, which is a satisfactory scenario~\cite{Becirevic:2017jtw}. Furthermore, this LQ contributes to the transition $b\to c\ell \bar \nu_{\ell^\prime}$, via the effective coupling,
\begin{align}\label{eq:gR2}
g_{S_L} = 4 g_T = \dfrac{v^2}{4 V_{cb}}\dfrac{y_L^{c\ell^\prime}\big{(}y_R^{b\ell}\big{)}^\ast}{m_{R_2}^2}\,,
\end{align}
at $\mu=m_{R_2}$. It can therefore accommodate the observed excess in $R_D$ and $R_{D^\ast}$, provided a large complex phase is present, cf.~Fig.~\ref{fig:rd-rdst-fit}.

\end{itemize}

\subsection{Vector leptoquarks}

\begin{itemize}
 \item[$\bullet$] $U_1=(\mathbf{{3}},\mathbf{1},2/3)$\,: A scenario with a weak singlet vector LQ attracted a lot of attention in the literature 
 since it provides the operators needed to explain both the $b\to c\tau\bar{\nu}$ and $b\to s\mu\mu$ anomalies~\cite{Calibbi:2015kma,Buttazzo:2017ixm,Kumar:2018kmr}.  
 The corresponding interaction Lagrangian can be written as 
\begin{equation}
\label{eq:U1}
 \begin{aligned}
\quad\;\;\; \mathcal{L}_{U_1} = x_L^{ij}\, \overline{Q}_i \gamma_\mu L_j\, U_1^\mu + x_{R}^{ij}\, \overline{d}_{R_i}\gamma_\mu \ell_{Rj} U_1^\mu+\mathrm{h.c.}\,,
 \end{aligned}
\end{equation}
where $x_{L}$ and $x_R$ stand for the $U_1$ couplings to fermions. Notice that the diquark couplings are absent for this state so that no additional assumption is needed. 
In its minimal setup, in which $x_R=0$, and starting from Eq.~\eqref{eq:U1}, one can easily obtain the contribution to $b\to s \ell_l^- \ell_k^+$,
\begin{equation}
 \begin{aligned}
  \delta C_9^{kl} = - \delta C_{10}^{kl} = - \dfrac{\pi v^2}{V_{tb} V_{ts}^\ast \alpha_\mathrm{em}} \dfrac{x_L^{sk} \big{(}x_L^{bl}\big{)}^\ast}{m_{U_1}^2}\,,
 \end{aligned}
\end{equation}
while for the $b\to c\ell\bar \nu_{\ell^\prime}$ one gets,
\begin{align}
g_{V_L} &=\dfrac{v^2}{2 V_{cb}}\dfrac{\big{(}V x_L\big{)}_{c\ell^\prime}\big{(}x_L^{b\ell}\big{)}^\ast}{m_{U_1}^2}\,.
\end{align}
In other words, this state alone can simultaneously explain $R_{K^{(\ast)}}$ and $R_{D^{(\ast)}}$, even in the minimal setup. 
The main reason for that to be the case is the absence of the tree level constraint coming from $\mathcal{B}(B\to K^{(\ast)}\nu\bar{\nu})$.

The challenge for extensions of the SM by a single vector LQ arises at the loop level because this scenario is non-renormalizable, 
which then undermines its predictiveness unless the ultraviolet (UV) completion is explicitly specified~\cite{Barbieri:2015yvd}. 
Several such completions have been proposed in the literature and they in general involve a $Z^\prime$ and a color-octet of vector bosons, 
in addition to the $U_1$ LQ itself, at the $\mathcal{O}(1\,\mathrm{TeV})$ scale~\cite{DiLuzio:2017vat}. 
In such situations additional assumptions on the spectrum of these states and on their couplings are required,  
which is a departure from the minimalistic scenarios described in this paper.

 \item[$\bullet$] $U_3=(\mathbf{{3}},\mathbf{3},2/3)$\,: Finally, the interaction of the weak triplet LQ with quarks and leptons is described by
\begin{equation}
 \begin{aligned}
\qquad\mathcal{L}_{U_3} = x_L^{ij}\, \overline{Q}_i \gamma_\mu  \big{(}\vec{\tau}\cdot \vec{U}_3^\mu\big{)} L_j +\mathrm{h.c.}\,,
 \end{aligned}
\end{equation}
where, as before, $x_{L}$ stands for the couplings to fermions. In contrast to $U_1$ this LQ allows for the dangerous diquark couplings, neglected in the Lagrangian above in order to ensure the proton stability. This scenario contributes to $b\to s \ell_l^-\ell_l^+$ via,
\begin{equation}
\delta C_9^{kl} = - \delta C_{10}^{kl} = -\dfrac{\pi v^2}{V_{tb} V_{ts}^\ast \alpha_\mathrm{em}} \dfrac{x_L^{sk} \big{(}x_L^{bl}\big{)}^\ast}{m_{U_3}^2}\,,
\end{equation}
which, again, can explain $R_{K}$ and $R_{K^\ast}$~\cite{Fajfer:2015ycq}, but it contributes to $b\to c\ell \bar \nu_{\ell^\prime}$ through
\begin{align}
g_{V_L} &=-\dfrac{v^2}{2 V_{cb}}\dfrac{\big{(}V x_L\big{)}_{c\ell^\prime}\big{(}x_L^{b\ell}\big{)}^\ast}{m_{U_3}^2}.
\end{align}
which is negative and therefore cannot accommodate  $R_{D}$ and $R_{D^\ast}$~\cite{Angelescu:2018tyl}, see Table~\ref{tab:bctaunu-fit}.
Furthermore, being a vector LQ, just like in the case of $U_1$, in this case too it is essential to specify the UV completion in order to remain predictive at the loop level. 
\end{itemize}

\section{LHC constraints}\label{sec:lhc}

Search for LQs in hadron colliders, either via their direct production~\cite{Diaz:2017lit,Dorsner:2018ynv} or through a study of the high-$p_T$ tails of the $pp\to \ell\ell$ distributions~\cite{Eboli:1987vb,Faroughy:2016osc,Angelescu:2020uug}, results in powerful constraints on the LQ masses and on their couplings to quarks and leptons. 
We provided such constraints in our previous paper~\cite{Angelescu:2018tyl}, which we update in the following by relying on the most recent LHC data.

\begin{table}[t!]
\renewcommand{\arraystretch}{1.85}
\centering
\begin{tabular}{|c|c|c|c|c|}
\hline 
Decays & Scalar LQ limits & Vector LQ limits&  $\mathcal{L}_\mathrm{int}$ / Ref. \\\hline\hline
 $ j j\,\tau\bar\tau$      & -- & -- & -- \\ 
 $ b\bar b\,\tau\bar\tau$  & $1.0~(0.8)$~TeV & $1.5~(1.3)$~TeV & $36~\mathrm{fb}^{-1}$~\cite{Aaboud:2019bye} \\ 
 $ t\bar t\,\tau\bar\tau$   & $1.4~(1.2)$~TeV & $2.0~(1.8)$~TeV & $140~\mathrm{fb}^{-1}$~\cite{Aad:2021rrh} \\ \hline
 $ j j\,\mu\bar\mu$     & $1.7~(1.4)$~TeV & $2.3~(2.1)$~TeV & $140~\mathrm{fb}^{-1}$ \cite{Aad:2020iuy}\\ 
 $ b\bar b\,\mu\bar\mu$   & $1.7~(1.5)$~TeV & $2.3~(2.1)$~TeV & $140~\mathrm{fb}^{-1}$ \cite{Aad:2020iuy} \\  
 $ t\bar t\,\mu\bar\mu$  &  $1.5~(1.3)$~TeV & $2.0~(1.8)$~TeV  & $140~\mathrm{fb}^{-1}$~\cite{Aad:2020jmj}\\  \hline
 $ j j\,\nu\bar\nu$       & $1.0~(0.6)$~TeV  & $1.8~(1.5)$~TeV & $36~\mathrm{fb}^{-1}$~\cite{CMS:2018bhq} \\ 
 $ b\bar b\,\nu\bar\nu$  & $1.1~(0.8)$~TeV & $1.8~(1.5)$~TeV & $36~\mathrm{fb}^{-1}$ \cite{CMS:2018bhq} \\ 
 $ t\bar t\,\nu\bar\nu$    & $1.2~(0.9)$~TeV & $1.8~(1.6)$~TeV & $140~\mathrm{fb}^{-1}$~\cite{Aad:2020sgw} \\ 
    \hline
\end{tabular}
\caption{ \sl \small Summary of the current limits from searches for pair-produced LQs at the LHC for possible final states (first column). Limits on scalar and vector LQs are shown in the second and third column, respectively, for a branching fraction $\beta=1$ ($\beta=0.5$). }
\label{tab:LQ-pair-bounds} 
\end{table}

\begin{figure*}[t!]
\centering
\includegraphics[width=.45\linewidth]{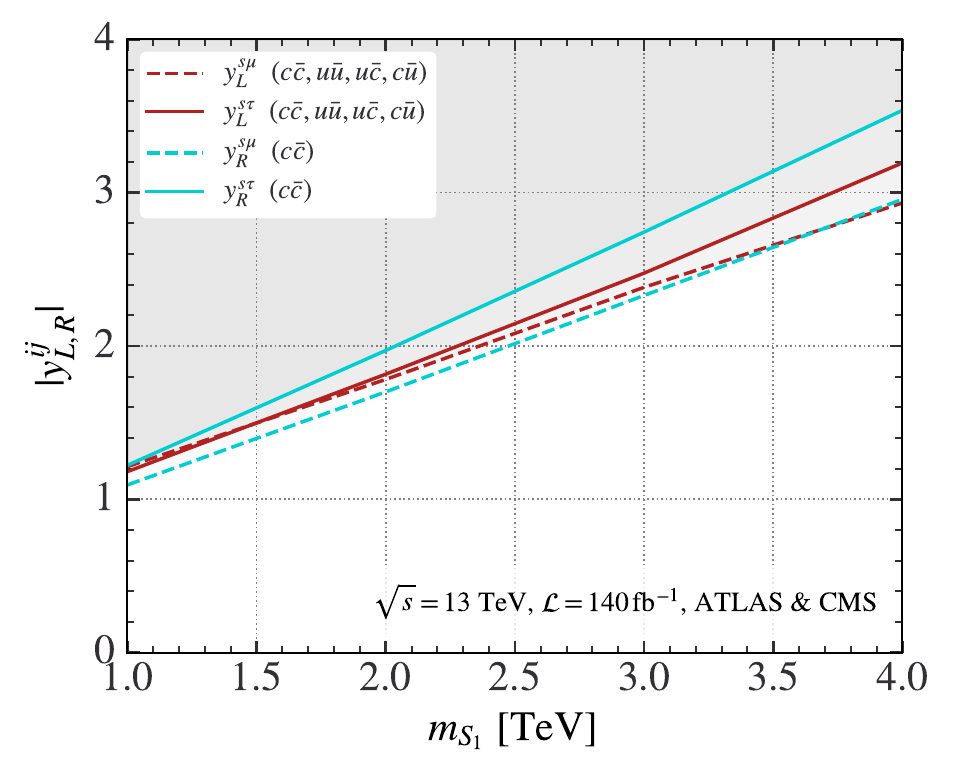}~\includegraphics[width=.45\linewidth]{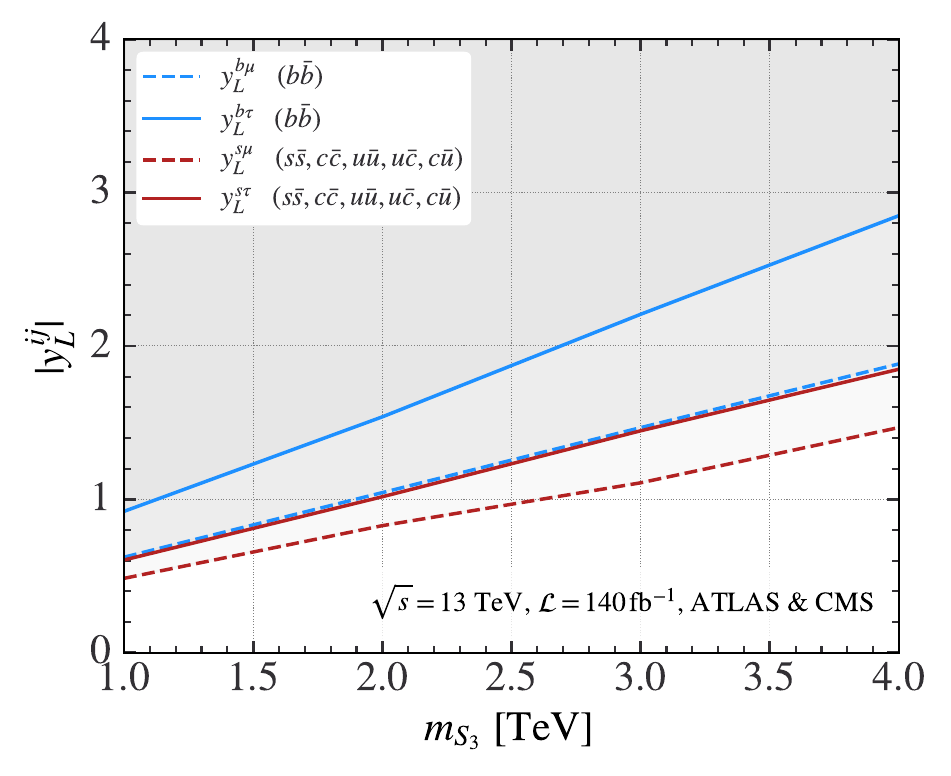}\\
\includegraphics[width=.45\linewidth]{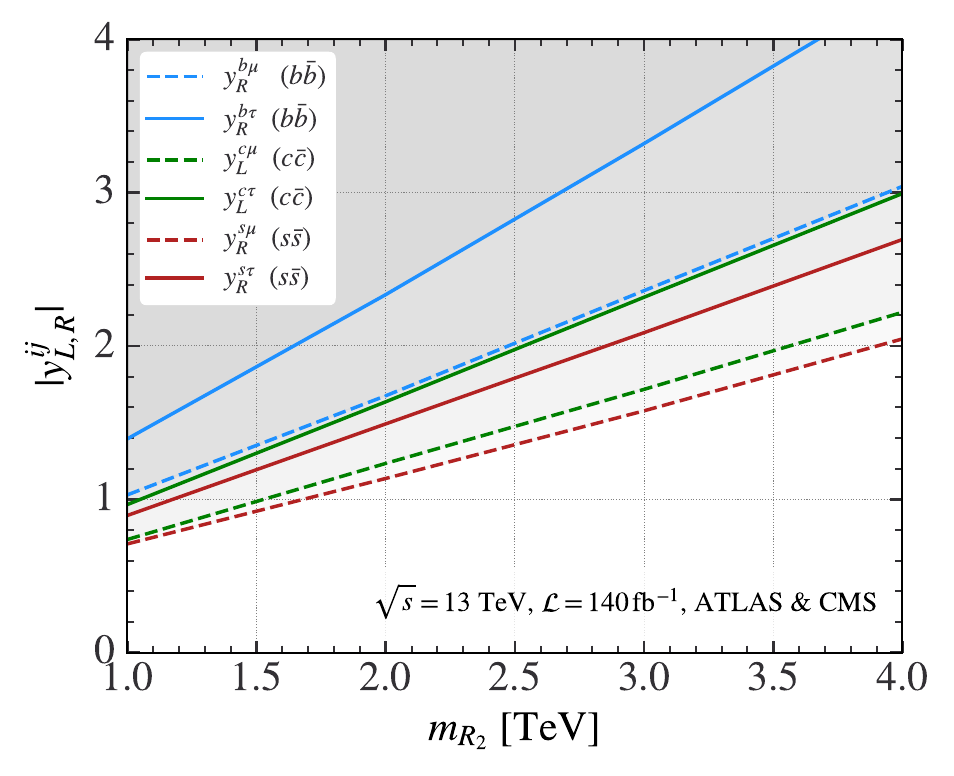}~\includegraphics[width=.45\linewidth]{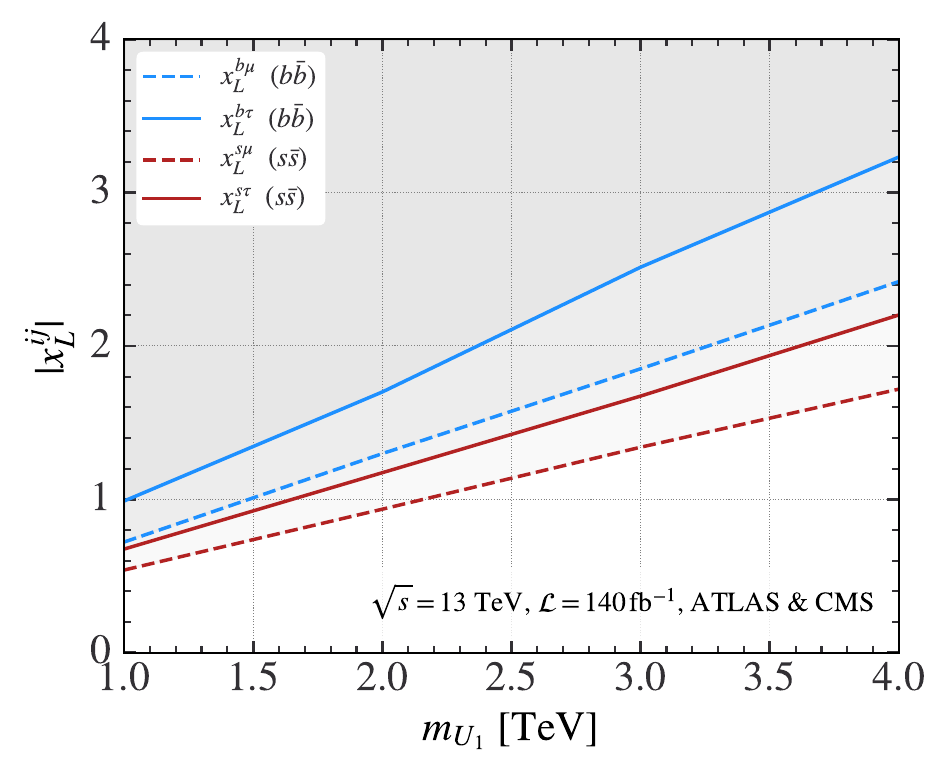}\\
\caption{\small \sl Upper limits on the scalar (vector) LQ couplings $y_L^{ij}$ ($x_L^{ij}$), as a function of the LQ masses, which have been obtained from the most recent LHC searches in the high-$p_T$ bins of $pp\to\ell\ell$ at 13~TeV with $140~\mathrm{fb}^{-1}$~\cite{Aad:2020zxo,CMS:2019tbu}. The solid (dashed) lines represent limits arising from di-muon (di-tau) searches, by turning on a single LQ coupling in flavor space at a time. The $q\bar q$ pairs inside the parentheses indicate the combination of $q\bar q\to\ell\ell$ channels used to set the exclusion limits for each coupling. Notice that all $u\bar u$ transitions are Cabibbo suppressed.} 
\label{fig:high-pT-plot}
\end{figure*}

\subsection{Direct searches}

The dominant mechanism for the LQ production at the LHC is $pp\to \mathrm{LQ}^\dagger\,\mathrm{LQ}$. Several searches for LQ pairs have been made at ATLAS and CMS for different final states, namely $(\bar{q}\ell)(q\bar{\ell})$, $(\bar{q}\nu)(q\bar{\nu})$ and $(\bar{q}_d\ell)(q_u\bar{\nu})$, where $q_d$ and $q_u$ stand for the generic down- and up-type quarks. From these searches it is possible to derive model independent bounds on a given LQ mass as a function of its branching fraction into a specific quark-lepton final state.

In Table~\ref{tab:LQ-pair-bounds} we present the new limits on the LQ masses obtained from our recast of the $pp\to \mathrm{LQ}^\dagger\,\mathrm{LQ}\to (\bar{q}\ell)(q\bar{\ell})$ ATLAS and CMS searches. These limits are obtained as a function of the LQ branching fraction $\beta$, which we take to the benchmark values $\beta = 1$ and $\beta = 0.5$. Our main assumption is that the LQ production cross-section is dominated by QCD, which is true for the range of Yukawa couplings allowed by flavor constraints~\cite{Angelescu:2018tyl}. Furthermore, we assume that the vector LQ ($V^\mu$) interaction with gluons ($G^{\mu\nu}$) is described by $\mathcal{L}\supset \kappa\, g_s V_\mu^\dagger G^{\mu\nu} V_\nu$, with $\kappa=1$ (Yang-Mills case)~\cite{Blumlein:1996qp}, and we use the predictions from \cite{Dorsner:2018ynv} in our recast. Note that the limits on LQs given in Table~\ref{tab:LQ-pair-bounds} are considerable improvements since our previous study~\cite{Angelescu:2018tyl}, thanks to $140~\mathrm{fb}^{-1}$ of the LHC data. As a result, we see that the overall lower limits on the LQ masses have been increased.

The LHC searches considered in Table~\ref{tab:LQ-pair-bounds} assume that pairs of LQs are produced and decay into the same quark-lepton final states. Recently, CMS performed a search for pair of LQs in the mixed channel $pp\to \mathrm{LQ}^\dagger \mathrm{LQ} \to  b \tau t \nu$, with $140~\mathrm{fb}^{-1}$ data~\cite{Sirunyan:2020zbk}. This search was performed under the assumption that the LQs decay with equal branching fractions ($\beta=0.5$) to the final states  $\mathrm{LQ}^{(2/3)}\to b \bar{\tau},\,t\bar{\nu}$, or  $\mathrm{LQ}^{(-1/3)}\to t\tau\,,b \nu$, where the upper index denotes the LQ electric charge. Under this assumption the lower limits $1.0$~TeV and $1.8~\mathrm{TeV}$ have been obtained for the scalar and vector LQs, respectively. That search is particularly useful for the $U_1 =(\mathbf{3},\mathbf{1},2/3)$ scenario, since the gauge invariance requirement implies that the couplings of $U_1$ to $t\bar{\nu}$ and to $b\bar{\tau}$ are equal. Note, however, that this search is very model dependent and, in particular, it does not generically apply to the models containing e.g.~$S_1=(\mathbf{\bar{3}},\mathbf{1},1/3)$ or $R_2=(\mathbf{{3}},\mathbf{2},7/6)$.

\subsection{Bounds from indirect high-$p_T$ searches}

Since the pioneering paper of Ref.~\cite{Eboli:1987vb} it is known that the high-energy tails of the invariant mass distribution of the processes $pp\to\ell\ell^{(\prime)}$~\cite{Faroughy:2016osc,Angelescu:2020uug} and $pp\to\ell\nu$~\cite{Greljo:2018tzh} are ideal probes for generic LQ models. These observables are particularly useful for setting upper bounds on complementary combinations of the couplings that cannot be constrained by flavor observables at low energies. In order to constrain the LQ couplings using LHC data, we follow a similar recasting procedure as outlined in Ref.~\cite{Angelescu:2018tyl}. The most recent ATLAS and CMS searches for resonances in the dilepton channels used here are:

\begin{itemize}
 \item[$\bullet$] $pp\to \tau^+\tau^-$: We recast the ATLAS search for heavy Higgs boson decaying into the $\tau\tau$ channel, at $\sqrt{s}=13~\mathrm{TeV}$ with $140~\mathrm{fb}^{-1}$ data \cite{Aad:2020zxo}. We consider events with hadronic $\tau$-leptons ($\tau_\mathrm{had}$) and we focus our analysis on the $b$-veto category.
 \item[$\bullet$] $pp\to \mu^+\mu^-$: We recast the CMS search for a heavy $Z^\prime$ boson decaying into the $\mu\mu$ channel, at $\sqrt{s}=13~\mathrm{TeV}$ with $140~\mathrm{fb}^{-1}$ data \cite{CMS:2019tbu}
 \end{itemize}

\noindent We do not recast LHC searches in the $pp\to \tau\nu$  mode since they are still only available with $36~\mathrm{fb}^{-1}$ data~\cite{Sirunyan:2018lbg,Aaboud:2018vgh}. Note, in particular, that gauge invariance under $SU(2)_L$ implies that large LQ contributions to $pp\to\ell\nu$ would necessarily appear in $pp\to\ell\ell$, which we consider in our study. Moreover, we do not recast the lepton flavor violating (LFV) modes such as $p p\to \ell\ell^\prime$, with $\ell\neq \ell^\prime$, since these constraints, in the specific case of LQs, turn out to be weaker than the combination of constraints arising from $p p\to \ell\ell$ and $p p\to \ell^\prime\ell^\prime$~\cite{Angelescu:2018tyl,Angelescu:2020uug}.\\

In this letter, we have refined the procedure for extracting our LQ limits in comparison to our previous paper~\cite{Angelescu:2018tyl}. The main differences are the following ones: 
\begin{itemize}
\item We perform a more conservative statistical analysis by using the so-called $\mathrm{CL}_s$ method~\cite{Read:2002hq}. The $95\%$ confidence level (CL) upper limits on the LQ couplings are obtained by profiling the likelihood ratio with the $q_\mu$ test statistics described in \cite{Cowan:2010js} and implemented  in the {\tt pyhf} package~\cite{Heinrich:2019}. Notice that the limits extracted using the CL$_s$ method  are much more resilient to possible statistical fluctuations in the experimental data populating low sensitivity regions of the spectrum, like e.g. the tails of the invariant mass. The resulting exclusion limits are therefore weaker when compared to the statistical method employed in~\cite{Angelescu:2018tyl}. Moreover, when performing the statistical analysis we have included a $20\%$ systematic uncertainty on the LQ signal. 

\item We take into account the interference of the $t$-channel LQ with the SM Drell-Yan process. Once included, these interference effects can have a moderate impact on the resulting limits, depending on the production channel. In particular, the constructive/destructive interference patterns can strengthen/weaken the naive limits from the $|\mathcal{A}_{\mathrm{NP}}|^2$ term up to $\mathcal{O}(20\%)$. 
\item Instead of showing limits from each individual $q\bar q\to  \ell\ell$ processes at a time, we provide limits for the individual couplings coming from different production channels. This results in more useful limits on the LQ couplings since they take into account all contributions, including the CKM-suppressed processes. For instance, the limits on the coupling $y_{L}^{s\ell}$ for the $S_3$ leptoquark are extracted from combining $s\bar s\to\ell\ell$, $c\bar c\to\ell\ell$, and the Cabibbo suppressed processes $u\bar u,u\bar c, c\bar u \to\ell\ell$.   
\item Our limits are also projected to the high-luminosity LHC phase with $3~\mathrm{ab}^{-1}$ in Sec.~\ref{sec:which}. To this purpose, we assume that the signal and background samples scale with the luminosity ratio, whereas all uncertainties scale with its square root. Although this assumption might appear too optimistic, it is worth stressing that higher $m_{\ell\ell}^2$ bins will become available with more data. Those higher bins are more sensitive to the LQ contributions than the bins that have been considered in the searches performed so far~\cite{Aad:2020zxo,CMS:2019tbu}.
\end{itemize}

Our constraints are collected in Fig.~\ref{fig:high-pT-plot} for the LQ models that are relevant for the $B$-physics anomalies, namely the scalars $S_1$, $S_3$ and $R_2$, and the vector $U_1$. In these plots we only present limits for the vector LQ couplings to left-handed currents.~\footnote{See Refs.~\cite{Baker:2019sli, Cornella:2021} for recent and updated high-$p_T$ limits for right-handed couplings.} The $95\%$ upper limits on the couplings are obtained as a function of the LQ masses by turning on one single flavor coupling at a time.  The specific $q\bar q\to\ell\ell$ transitions contributing to each exclusion limit are displayed inside the parentheses $(q\bar q)$. As shown in Fig.~\ref{fig:high-pT-plot}, these limits are typically more stringent than naive perturbative bounds on the couplings, namely $|y|\lesssim \sqrt{4\pi}$. The relevance of these constraints to the scenarios aiming to explain $R_{K^{(\ast)}}$ and $R_{D^{(\ast)}}$ will be discussed in Sec.~\ref{sec:which}.

\section{Which leptoquark?}\label{sec:which}

In Table~\ref{tab:LQ-lists} we summarize the situation regarding the viability of a scenario in which the SM is extended by a single 
$\mathcal{O}(1\,\mathrm{TeV})$ LQ state. We now comment and provide useful information for each one of them.
\begin{table}[t!]
\renewcommand{\arraystretch}{1.9}
\centering
\begin{tabular}{|c|cc||c|}
\hline 
Model &  ~$R_{K^{(\ast)}}$~ & ~$R_{D^{(\ast)}}$~ & $R_{K^{(\ast)}}$ $\&$ $R_{D^{(\ast)}}$\\ \hline\hline
~$S_3$~~$(\mathbf{\bar{3}},\mathbf{3},1/3)$	& $\color{blue}\checkmark$	&\color{red}\xmark	 &\color{red}\xmark	\\   
~$S_1$~~$(\mathbf{\bar{3}},\mathbf{1},1/3)$	& \color{red}\xmark & $\color{blue}\checkmark$	& \color{red}\xmark	\\ 
~$R_2$~~$(\mathbf{{3}},\mathbf{2},7/6)$ &	\color{red}\xmark & $\color{blue}\checkmark$	&\color{red}\xmark\\ \hline
~$U_1$~~$(\mathbf{{3}},\mathbf{1},2/3)$ 	& $\color{blue}\checkmark$	& $\color{blue}\checkmark$	& $\color{blue}\checkmark$	\\
~$U_3$~~$(\mathbf{{3}},\mathbf{3},2/3)$ & $\color{blue}\checkmark$	&\color{red}\xmark	&\color{red}\xmark	\\  \hline
\end{tabular}
\caption{ \sl \small Summary of the LQ models which can accommodate $R_{K^{(\ast)}}$ (first column), $R_{D^{(\ast)}}$ (second column), and both $R_{K^{(\ast)}}$ and $R_{D^{(\ast)}}$ (third column), without being in conflict with existing constraints. See text for details.}
\label{tab:LQ-lists} 
\end{table}
\begin{itemize}
 \item[$\circ$] $S_3$: With respect to our previous paper, the situation in the scenario with a triplet of mass degenerate scalar LQs did not significantly change. This scenario is indeed the best scalar LQ solution to describing the current $B$-physics anomaly $R_{K^{(\ast)}}^\mathrm{exp}<R_{K^{(\ast)}}^\mathrm{SM}$, which is why it is often combined in the literature with another scalar LQ so as to accommodate both  $R_{K^{(\ast)}}^\mathrm{exp}<R_{K^{(\ast)}}^\mathrm{SM}$ and  $R_{D^{(\ast)}}^\mathrm{exp}>R_{D^{(\ast)}}^\mathrm{SM}$.

\item[$\circ$] $S_1$: As noted in Eq.~\eqref{eq:gVSS1}, even in the minimalistic scenario (with $y_R^{ij} =0$), $S_1$ alone can reproduce the observation $R_{D^{(\ast)}}^\mathrm{exp}>R_{D^{(\ast)}}^\mathrm{SM}$. In the non-minimal case ($y_R^{ij} \neq 0$), the additional coupling, $g_{S_L}=-4\, g_T$, also provides a viable solution to this problem, cf.~Fig.~\ref{fig:rd-rdst-fit}. This scenario, however, does not lead to a desired contribution to the $b\to s\mu\mu$. In the minimal ansatz for the Yukawa couplings accommodating  $R_{K^{(\ast)}}^\mathrm{exp}<R_{K^{(\ast)}}^\mathrm{SM}$ and $\Delta m_{B_s}$ requires large LQ mass, $m_{S_1} \gtrsim 4$~TeV, and at least one of the Yukawa couplings to hit the perturbativity limit $\sqrt{4\pi}$~\cite{Angelescu:2018tyl}. 
Therefore, one needs to turn on at least $y_R^{c\tau}$ and otherwise satisfy the condition $|y_{R}^{i\mu}|\ll |y_{L}^{i\mu}|$, for $i \in \lbrace u,c,t \rbrace$ to be consistent with data, cf. Fig.~\eqref{fig:c9-c10-fit}. However, requiring consistency with a number of measured flavor physics observables~\cite{Angelescu:2018tyl}, including $R_{D^{(\ast)}}^{\mu/e} = \mathcal{B}(B\to D^{(\ast)} \mu\bar{\nu})/\mathcal{B}(B\to D^{(\ast)} e \bar{\nu})$, $\mathcal{B}(B\to K^{(\ast)}\nu\bar{\nu})$, $\mathcal{B}(K\to \mu \nu)/\mathcal{B}(K\to e \nu)$ and the experimental limit on $\mathcal{B}(\tau \to \mu \gamma)$, leads to a large $m_{S_1}$ and very large couplings. This is why the $S_1$ scenario is considered as unacceptable for describing $R_{K^{(\ast)}}^\mathrm{exp}<R_{K^{(\ast)}}^\mathrm{SM}$, but fully acceptable for describing $R_{D^{(\ast)}}^\mathrm{exp}>R_{D^{(\ast)}}^\mathrm{SM}$. cf. Refs.~\cite{Angelescu:2018tyl,Becirevic:2016oho,Cai:2017wry}.

\item[$\circ$]$R_2$: Clearly, on the basis of Eq.~\eqref{eq:gR2} and the results presented in Table~\ref{tab:bctaunu-fit} and Fig.~\ref{fig:rd-rdst-fit}, this scenario can be viable for enclosing $R_{D^{(\ast)}}^\mathrm{exp}>R_{D^{(\ast)}}^\mathrm{SM}$, if at least one $y_R^{ij}$ is non-zero, usually $y_R^{b\tau}$. 
In fact, it suffices to allow $y_{L}^{c\tau} \,\big{(}y_R^{b\tau}\big{)}^\ast$ to be $\mathcal{O}(1)$ to ensure the compatibility both with the low-energy observables and with direct searches at LHC, as shown in Fig.~\ref{fig:high-pT-plot}. As mentioned before, this LQ scenario generates the combination $g_{S_L} = 4\, g_T$ at the matching scale $\mu\simeq m_{R_2}$, which is consistent with data if $g_{S_L}$ is mostly imaginary, cf.~Fig.~\ref{fig:rd-rdst-fit} and Refs.~\cite{Becirevic:2018uab,Sakaki:2013bfa,Hiller:2016kry}. 

Like in the $S_1$ scenario, this LQ cannot generate the tree level contribution consistent with $R_{K^{(\ast)}}< R_{K^{(\ast)}}^{\mathrm{SM}}$, but it can do so through the box-diagrams~\cite{Becirevic:2017jtw}. The two essential couplings for this to be the case, $y_L^{c\mu}$ and $y_L^{t\mu}$, can now be quantitatively scrutinized. To that end it is enough to  use two key constraints: the one arising from the well measured $\mathcal{B}(Z\to \mu\mu)$~\cite{Zyla:2020zbs} and another one, stemming from the high-$p_T$ tail of the $pp\to \mu\mu$ differential cross section. 
Note that the expression for the corresponding LQ contribution to $Z\to \mu\mu$ has been recently derived in Ref.~\cite{Arnan:2019olv}, where the non-negligible finite terms $\propto x_Z\log x_t$ have been properly accounted for ($x_i=m_i^2/m_{R_2}^2$). As for the LQ mass, we use the bound given in Table~\ref{tab:LQ-pair-bounds} and set $m_{R_2}=1.7$~TeV, while  from Fig.~\ref{fig:high-pT-plot} we can read off the constraints on the couplings as obtained from the large $p_T$ considerations. The result is shown in Fig.~\ref{fig:R2-plot} where we also draw the curves corresponding to three significant values of $R_{K^{(\ast)}}$, making it obvious that only 
$R_{K^{(\ast)}}\gtrsim 0.9$ is compatible with the two mentioned constraints. 
In other words, $R_{K^{(\ast)}}$ in this scenario is pushed to the edge of $1\sigma$ compatibility with $R_{K^{(\ast)}}^\mathrm{(exp)}$, cf. also Ref.~\cite{Camargo-Molina:2018cwu}. 
 
\begin{figure}[t!]
\centering
\includegraphics[width=1.\linewidth]{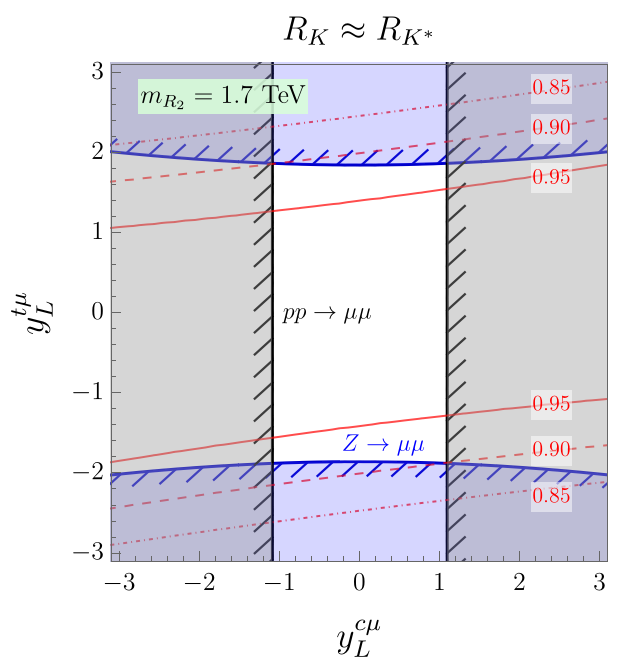}
\caption{\small \sl The allowed regions for the couplings $y_L^{c\mu}$ and $y_L^{t\mu}$ are plotted in white for the $R_2=(\mathbf{3},\mathbf{2},7/6)$ LQ with mass $m_{R_2}=1.7$~TeV. Predictions for $R_{K}\approx R_{K^\ast}$ in the bin $q^2\in [1,6]~\mathrm{GeV}^2$ are shown by the red contours. Excluded regions by $Z$-pole observables and $pp\to\mu\mu$ constraints are depicted in blue and gray, respectively.} 
\label{fig:R2-plot}
\end{figure}

As discussed in our previous paper, the simultaneous explanation of both $R_{K^{(\ast)}}$ and $R_{D^{(\ast)}}$ in this scenario is not possible even to $2\sigma$ because of the chiral enhancement by the top quark which leads to a prohibitively large $\mathcal{B}(\tau \to \mu \gamma)$, in conflict with the experimental bound~\cite{Becirevic:2017jtw}.

\begin{figure*}[t!] 
\centering
\includegraphics[width=.5\linewidth]{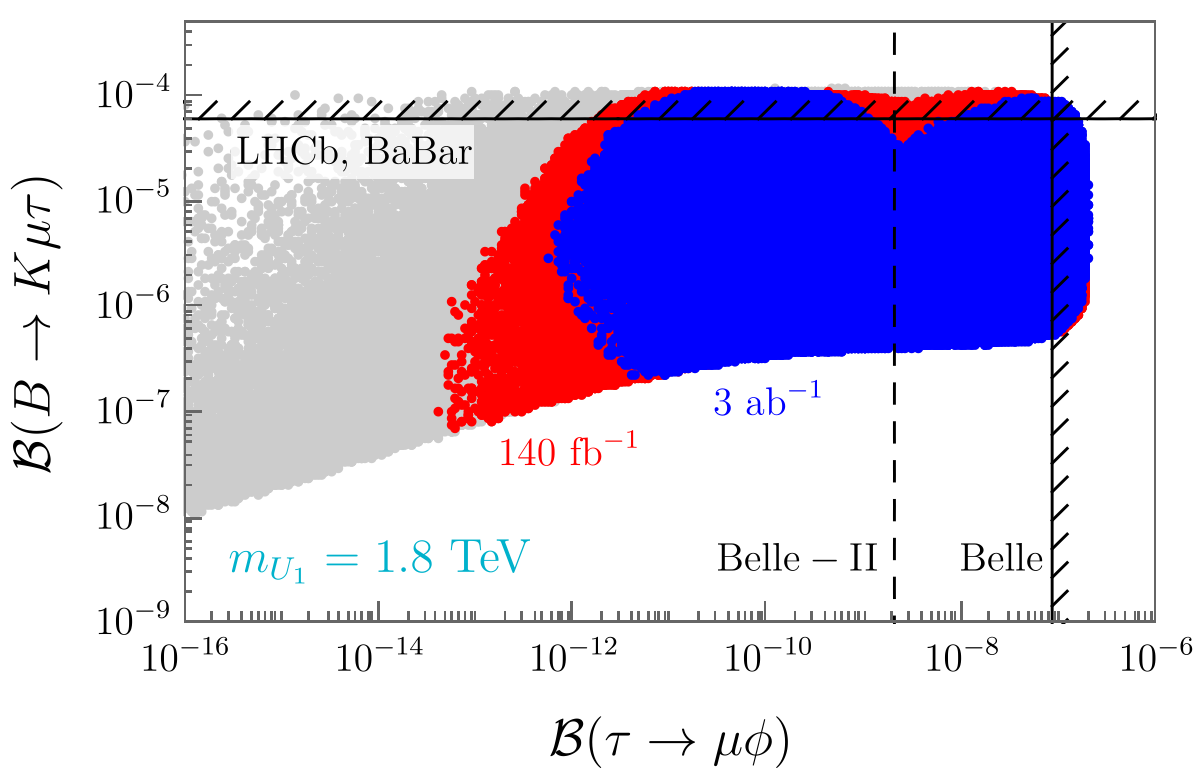}~\includegraphics[width=.489\linewidth]{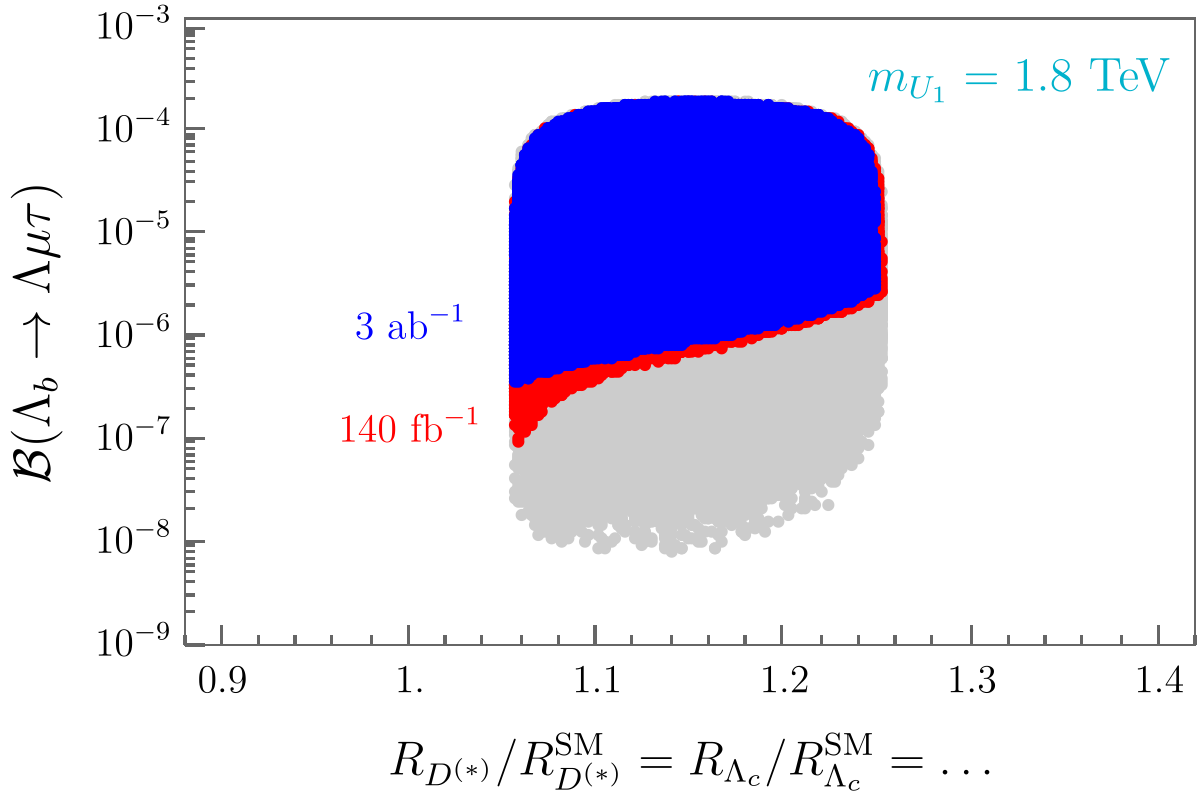}
\caption{\small \sl Lower and upper bounds on the exclusive $b\to s \mu\tau$ processes as obtained in the minimal $U_1$ scenario from the constraints arising both from the low-energy observables (gray points) and those coming from the current direct searches at the LHC (red points), the subset of which (blue points) correspond to the projected integrated luminosity of $3~\mathrm{ab}^{-1}$.  }  
\label{fig:U1-plot}    
\end{figure*} 

 \item[$\circ$] $U_1$: Owing to the fact that this LQ does not contribute to $B\to K^{(\ast)}\nu\bar{\nu}$ at tree level, this is the only scenario that can satisfy both anomalies. 
The main drawback, however, is that the constraints derived from the loop induced processes cannot be used unless a clear UV completion is specified which in turn requires introducing several new parameters and new assumptions (model dependence) making the scenario less predictive. 
In our previous paper~\cite{Angelescu:2018tyl} we made a detailed analysis and found that this scenario can be significantly constrained by the tree level processes alone, cf. also Ref.~\cite{Cornella:2019hct}. In particular we showed that the model results in interesting correlation between the LFV processes $B\to K^{(\ast)} \mu\tau$ and $\tau\to \mu \phi $, and both the upper and lower bounds for these modes have been derived. 
With respect to our previous paper, the lower bound on $m_{U_1}$ has increased and we set it to $m_{U_1} =1.8$~TeV, see Table~\ref{tab:LQ-pair-bounds}. We then use the low energy flavor physics observables as in Ref.~\cite{Angelescu:2018tyl}, combine them with the new constraints on couplings, as obtain from the high-$p_T$ shapes of $pp\to \ell\ell$, shown in Fig.~\ref{fig:high-pT-plot}, and instead of plotting the couplings, we focus directly onto observables. Using the expressions for exclusive LFV $b\to s\ell_1\ell_2$ modes~\cite{Becirevic:2016oho,florentin2} in the first panel of Fig.~\ref{fig:U1-plot} we show how the region of $\mathcal{B}(B\to K\mu\tau)$ and $\mathcal{B}(\tau \to \mu \phi )$, allowed by the low-energy flavor physics constraints (gray points), gets reduced to the red region, once the current constraints coming from the high $p_T$ considerations of $pp\to \ell\ell$ at the LHC are taken into account. We see that in both channels the current experimental bounds are already eliminating small sections of the parameter space. In the same plot we also show how that experimental bound on $\mathcal{B}(\tau \to \mu\phi  )$ is expected to be lowered once the Belle~II runs will be completed~\cite{Kou:2018nap}. Concerning the experimental bound on $\mathcal{B}(B\to K\mu\tau)$, we note that the BaBar bound ($4.8 \times 10^{-5}$)~\cite{Lees:2012zz} has been recently confirmed and slightly improved by LHCb ($3.9 \times 10^{-5}$)~\cite{Aaij:2020mqb}. In the minimal $U_1$ scenario considered here, and with the current experimental constraints, we obtain 
\bea\label{boundLFV2}
\mathcal{B}(B\to K \mu\tau) \gtrsim 0.7 \times 10^{-7}\,, 
\eea
which could be tested experimentally. Note that this (lower) bound is not expected to increase significantly with the improved luminosity of the LHC data, and with the projected $3~\mathrm{ab}^{-1}$ of data we get only a factor of about $3$ improvement, namely $\mathcal{B}(B\to K \mu\tau) \gtrsim 2.2 \times 10^{-7}$.

We should also mention that, in this scenario, from the lower bound~\eqref{boundLFV2} and the experimental upper bound, one can derive the bounds on similar decay modes since 
${\mathcal{B}(B\to K^\ast \mu\tau)/\mathcal{B}(B\to K\mu\tau)}\approx 1.8$, ${\mathcal{B}(B_s\to \mu\tau)/\mathcal{B}(B\to K\mu\tau)}\approx 0.9$, and 
 ${\mathcal{B}(\Lambda_b\to \Lambda \mu\tau)/\mathcal{B}(B\to K\mu\tau)}\approx 1.7$~\cite{florentin2}. 
Furthermore, in this scenario the SM contribution to the $b\to c\tau \bar\nu$ decay modes gets only modified by and overall factor. For that reason, the predicted increase 
of $R_X$ with respect to the SM is the same for any $X\in \{ D^{(\ast)}, D_s^{(\ast)}, J/\psi, \Lambda^{(\ast )}_c, \dots \}$.
From the right panel of Fig.~\ref{fig:U1-plot} we see that with the current experimental constraints we have  
\bea
1.05 \lesssim {R_X\over R_X^\mathrm{SM}} \lesssim 1.25 \,,
\eea
the interval which remains as such even by projecting to $3~\mathrm{ab}^{-1}$ of the LHC data (blue regions in Fig.~\ref{fig:U1-plot}).


\end{itemize}

\section{Conclusions}\label{sec:conclusions}

In this work we revisited our previous phenomenological study and examined the viability of the scenarios in which the SM is extended by only one 
$\mathcal{O}(1\,\mathrm{TeV})$ LQ after comparing them to the most recent experimental results, in addition to those already discussed in our Ref.~\cite{Angelescu:2018tyl}. 
In that respect the  Belle measurement of $R_{D^{(\ast )}}$~\cite{Abdesselam:2019dgh} has been particularly important, as well as the new $R_K$ and $\mathcal{B}(B_s\to \mu\mu)$ values reported by the LHCb Collaboration~\cite{1852846,LHCbNEW}.  Besides the low-energy observables, we also exploit the most recent experimental improvements regarding the direct searches and the high $p_T$ considerations of the $pp\to \ell\ell$ differential cross section studied at the LHC. 

Better experimental bounds on the LQ pair production, $pp\to \mathrm{LQ}^\dagger\,\mathrm{LQ}$, results in a larger lower bound on $m_\mathrm{LQ}$, now straddling $2$~TeV and being higher for the vector LQs than that for the scalar ones. From the study of the large-$p_T$ spectrum of the differential cross section of $pp\to \ell\ell$, we extract the upper bounds on Yukawa couplings which provide us with constraints complementary to those inferred from the low-energy observables.

Whenever available we use the improved theoretical expressions and improved hadronic inputs. On the basis of our results, which are summarized in Table~\ref{tab:LQ-lists},  
we confirm that none of the scalar LQs alone, with the mass $m_\mathrm{LQ}\lesssim 2$~TeV, can be a viable scenario of NP that captures both types of anomalies, $R_{K^{(\ast)}}^\mathrm{exp}<R_{K^{(\ast)}}^\mathrm{SM}$ and  $R_{D^{(\ast)}}^\mathrm{exp}>R_{D^{(\ast)}}^\mathrm{SM}$. Instead, one can combine $S_3$ with either $S_1$ or $R_2$~\cite{Becirevic:2018afm,Saad:2020ucl,Gherardi:2020qhc,Crivellin:2017zlb} to get a model suitable for describing all of the data in a scenario requiring the least number of parameters. 

With the new experimental data we were able to better examine the model with $R_2$ scalar LQ, and check on the possibility of describing the $R_{K^{(\ast)}}^\mathrm{exp}<R_{K^{(\ast)}}^\mathrm{SM}$
anomaly through the loop process. We found that $\mathcal{B}(Z\to \mu\mu)$ and the constraint coming from the high $p_T$ shape of the $pp\to \mu\mu$ cross section at the LHC are complementary to each other and allow us to rule out the model (to $1\sigma$) if $R_{K^{(\ast)}}\lesssim 0.9$.

Besides the scalar LQs we also considered the vector one, $U_1$, for which we could not account for the loop induced processes, such as $\Delta m_{B_s}$, but by focusing on the tree level observables alone we could confirm that this scenario, in its minimal setup ($x_R=0$) can describe both $R_{D^{(\ast)}}^\mathrm{exp}> R_{D^{(\ast)}}^\mathrm{SM}$ and $R_{K^{(\ast)}}^\mathrm{exp}< R_{K^{(\ast)}}^\mathrm{SM}$. 
In this $U_1$ model all the exclusive processes based on $b \to c\tau \bar \nu$ are modified by the same multiplicative factor so that all the LFUV ratios are the same. In other words, and with the currently available experimental information,   
$1.05 \lesssim R_X/R_X^\mathrm{SM} \lesssim 1.25$, $X\in\{ D^{(\ast)}, D_s^{(\ast)}, J/\psi,  \Lambda^{(\ast )}_c,\dots \}$.
Also interesting are the upper and lower bounds on the LFV $b\to s\mu\tau$ modes. While the upper bound is already superseded by the experimentally established one, this scenario provides us with the lower bound, which we found to be $\mathcal{B}(B\to K\mu\tau) \gtrsim 0.7 \times 10^{-7}$. In this study we also included baryons and obtain  $1.2\times 10^{-7} \lesssim \mathcal{B}(\Lambda_b \to \Lambda \mu\tau) \lesssim 6.6 \times 10^{-5}$, where the lower bound is a prediction of the $U_1$ model discussed here, and the upper bound is obtained by rescaling the experimental bound on $\mathcal{B}(B\to K\mu\tau)$. 
 
\vspace*{1.2em}

\section*{Acknowledgments}

This project has received support from the European Union’s Horizon 2020 research and innovation programme under the Marie Skłodowska-Curie grant agreement No 860881-HIDDeN. The work of D.A.F. has received funding from the European Research Council (ERC) under the European Union’s
Horizon 2020 research and innovation programme under grant agreement 833280 (FLAY), and by the Swiss National Science Foundation (SNF) under contract 200021-
175940.

\end{document}